# The "go-stop-go" of Italian civil nuclear programs, beset by lack of strategic planning, exploitation for personal gain and unscrupulous political conspiracies: 1946-1987

**Angelo Baracca [a], Giorgio Ferrari [b] and Roberto Renzetti [c]**

[a] Department of Physics, University of Florence; [b] Former ENEL specialist; [c] Physicist, www.fisicamente.net

**Angelo Baracca** (corresponding author)
Department of Physics and Astronomy, University of Florence
via G. Sansone, 1
 50019 Sesto Fiorentino (FI)
Tel.: (39)3280196987
e-mail: baracca@fi.infn.it

**Giorgio Ferrari**
Former Enel nuclear expert
Via A.Dulceri,171
00176 Roma
Tel.: (39)064513698
e-mail: giorgio.ferrari@autoproduzioni.net

**Roberto Renzetti**
Physicist www.fisicamente.net
Via Avegno 58
00165 Roma (RM)
Tel (39)0666417178
e-mail: fisicamente@fisicamente.net

### Abstract

The Italian civil nuclear projects had a very early origin, with the first ideas originating as far back as 1945. The construction of the first three plants dated back to the period 1956-1964, and at that time Italy ranked third in the world for installed power. However, the very ambitious projects were immediately after thwarted in the early 1960s by the "Ippolito trial". Actually, a whole range of advanced programmes for the development of the country went to a stop, since they clashed with national and international major powers. Italy was relegated to a second rank power. The fourth nuclear plan was designed in 1970, and its commercial operation began in 1981. In the meantime, a strong anti-nuclear movement grew, and the position of the pro-nuclear Italian Communist Party began to change. After the Chernobyl accident, a national referendum was held, which in 1987 put an end to the Italian nuclear programmes.

## 1. Introduction. The framework

   Nuclear programmes are always, in every country, very complex processes, due to the complexity of this technology and the huge economic and political interests it involves. But in Italy further complications, typical of the Italian system, usually arise in any process such as conflicts of interest, off-stage plots, and dark manoeuvres, if not corruption. The debate on the nuclear programmes in Italy has been extremely harsh. Popular movements have been very strong and polarized in certain phases, and the evaluations of these programmes by different social components have often been quite different, or divergent. The international context has played an important role in many phases of the Italian nuclear programmes, not always in transparent ways. For these reasons, it is not simple to reconstruct a true history of the Italian nuclear programmes.[1] Moreover, in our opinion a full reconstruction cannot be restricted to the analysis of official documents and archives, though these are of course very important, however deficient they may be. It must consider internal and external political events, and the evolution of power relations in the country, with dynamics and motivations that are often far from clear-- Italy is, after all, the country of mysteries. Some reconstructions reflect partisan views on nuclear power. [2]

While trying to take into account all these aspects, our reconstruction will necessarily be very concise.

## 2. Some peculiar features of the situation

There is a number of both internal and international factors that we consider significant, including typical drawbacks and contradictions of Italian affairs.

i) Italy came out of the war, socially and economically, deeply destroyed. There was no technologically advanced industry or know-how. In the first two post-war decades, the private electric companies were among the strongest economic and political powers.

ii) Before the war there was no scientific research centre, apart from the Institute of Physics in Rome, in which Enrico Fermi and his group carried out the fundamental neutron experiments in 1934: [3] but the Fermi group had no real possibility of acquiring the equipment necessary to develop its programmes, and the racial laws of 1938 were the final impulse for most members of the group to leave Italy, along with other physicists. Only Edoardo Amaldi stayed in Italy, trying to keep up an activity of fundamental research under the Nazi occupation,[4] and to rebuild and relaunch Italian physics after the war.[5] There was therefore a serious shortage of trained staff.

iii) In these precarious conditions, it appears noteworthy that the radical novelty and potential of nuclear energy were grasped immediately after the nuclear explosions on Japan by a small number of young physicists and engineers. Amaldi, and some other physicists, supported the early projects, perceiving the strong relationship between technology and fundamental physics, in which they were more interested. These young scholars had the capacity for establishing contacts with the right people, and translating this early perception into concrete projects in the field of nuclear technology.

iv) In spite of these precarious conditions, Italy was the only country among the war losers which successfully planned and undertook projects and activities in the field of nuclear energy. Germany was forbidden for a long time to develop activities in this field. Japan was a very different story, since the US carried out in that country a subtle and widespread propaganda campaign aimed at minimizing the events of Hiroshima and Nagasaki, exalting the progressive role of "civil" nuclear technology.[6]

v) However, the enlightened Italian promoters of nuclear projects faced at the beginning a crucial precondition, since the peace treaty under negotiation in Paris (with Italy in a marginal role) could have forbidden every activity in the applied nuclear field. Not before autumn 1946 was such a danger thwarted. But only the subscription by the country of the NATO treatise in 1949 gave the green light to a nuclear agreement with the US in 1955 (during the take off of the Atoms for Peace campaign).

vi) In this complex context, the scant group of young Italian promoters, with no direct experience in the new field, planned and actively started in the following couple of decades a wide spectrum of nuclear projects, covering almost the full range of the nuclear cycle, including the design and construction of a domestic nuclear reactor.

vii) The Italian political class and governments showed however great indifference, and ignorance, towards nuclear energy. The early initiatives in this field were undertaken by private companies (CISE, 1946). The earliest State initiative had to wait the year 1952 (CNRN). This situation contrasts sharply with what happened in the other countries, where specially-constituted public bodies were established (AEC in the US, CEA in France, AEA in the UK).

viii) However, the private initiative introduced a deep contradiction, since the powerful private electrical industry strenuously opposed the nationalization of the electric sector. Paradoxically, this contradictory situation brought to the surprisingly quick and precocious construction of the first three nuclear plants (1955-1964), among the earliest in the world. But this sudden start did not reflect a coherent plan, and took place in the absence of a national law for the nuclear sector, which was seen by private industry as a step towards nationalization.

ix) Just when this remarkable result was being achieved, in the early 1960s, the Italian nuclear ambitions came to an abrupt stop, due to the explosion of a blatant scandal, whose true causes and instigators have never been completely clarified. Actually, the nuclear choice was opposed by powerful, mainly international, forces that pushed for a subordinate role of the country in the Western block. In their view, Italy was to remain an energy-dependent producer of mass consumer goods, basically excluded from advanced technologies. After the stop to the nuclear programmes, Italy became completely dependent on oil (see Figure 1).

**Figure 1. Historical evolution of post-war production of electric energy in Italy.**

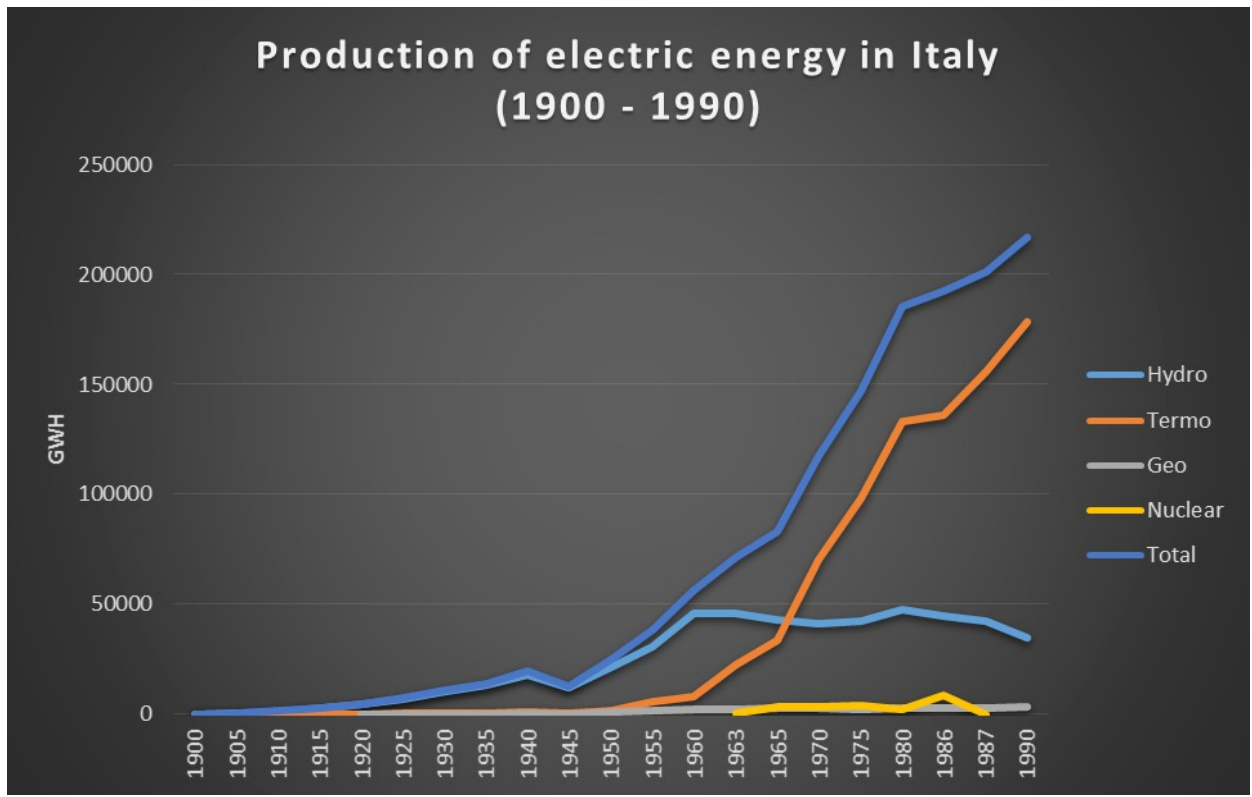

x) The Italian military have played a marginal role in the field of nuclear energy, although in some phases they have designed some ambitious projects.

### 3. Phases of development of the Italian nuclear programmes

The development of the Italian nuclear programmes may be divided into distinct phases, separated by sudden breaks, amid deep contradictions.

After the early private initiatives as far back as 1946, the sudden take-off between 1955 and the early 1960s projected Italy among the top nuclear countries in the world. However, this occurred without a coherent programme or a nuclear law, although the ambitious target was the installation of 6,000 MW of nuclear power by 1975, a figure never reached during the whole history of Italian nuclear programmes. There was even an original project of an Italian natural uranium, heavy water "fog" reactor, called "Cirene". This was never completed, and resulted in a great waste of money (see Table 1).

**Table 1. The four Italian power reactors**

| Place | Type | Power | Date start [b] | Date shut down |
|---|---|---|---|---|
| Latina [c] | Gas-graphite | 210 ᴍᴡᴇ | May 1963 | November 1986 |
| Garigliano | ʙᴡʀ General Electric | 150 ᴍᴡᴇ | January 1964 | August 1978 |
| Trino Vercellese | ᴘᴡʀ Westinghouse | 260 ᴍᴡᴇ | October 1965 | March 1987 |
| Caorso | ʙᴡʀ General Electric | 860 ᴍᴡᴇ | May 1978 | October 1986 |

[a] To these, one must add the *research reactors* Avogadro, Tapiro, Triga, Raptus, Exor (Ispra), the Cirene, the never finished fast reactor PEC, and the *military research reactor* at CAMEN near Pisa.

[b] date of connection to the electric grid.

[c] At the date of its start it was the most powerful nuclear reactor in Europe.

However, in 1963-64 these developments suffered a drastic standstill, when the General Secretary of the Italian Commission for Nuclear Energy (CNEN), Felice Ippolito, was charged and condemned for administrative irregularities. Only in 1971 the construction began of the bigger nuclear plant at Caorso, which however had to wait 1981 to start operating.

During the Seventies and Eighties a sequence of more or less pharaonic National Energy Programmes (PEN) was proposed by successive Governments, but strong popular protests were growing. Actually, the history of the Italian nuclear programmes cannot be considered complete until a full history of popular movements is reconstructed. And this is a difficult challenge, since popular movements hardly ever leave documentary traces of their organization and actions, so that we can only resort to oral history, or newspaper articles. In fact, while the construction of a fifth nuclear plant began, a popular referendum was launched, intended to shut down altogether the Italian nuclear programmes. The impression of the 1986 Chernobyl incident contributed to the result of the referendum in 1987, which put an end to Italian nuclear ambitions (but left behind a heavy inheritance, a real emergence which is far from being closed after three decades).

A quarter of a century later, the short-lived attempt at relaunching nuclear energy by the Berlusconi government in 2008 was buried after the Fukushima disaster by a second national referendum in 2011. In the meantime almost nothing has been done about the inheritance left by the previous programmes, leaving the problems of the decommissioning of four reactors and of the disposal of nuclear waste practically unsolved, with heavy economic burdens and environmental and health hazards.

## 4. The postwar economic and industrial situation. Big interests and energy choices

Although Italy suffered severe destruction from the war, damages to industrial structures were relatively limited. Thanks in part to the partisans and workers who occupied the factories, a network of recently built plants was left standing mainly in the industrialized North, although with a capacity utilization rate of only around 50%[7]. The collapse of German competition and cheap labour opened great prospects for the Italian industry.

The electric and steel industries had acquired a strong power under fascism and in wartime, thanks to their close connections with the regime: after the war they were in the best position to exploit the difficult situation of the country, and strongly opposed any policy of economic

planning.[8]

Reconstruction and industrial restructuring was based essentially on mature technologies, and the strong economic development of the years 1950-1963 owed much to the low cost of labour and labour-intensive production in the car industry. Electricity production was almost entirely of hydroelectric origin, and completely in the hands of highly concentrated private companies: in 1941, 8 firms out of the existing 320 controlled 77% of the electricity production; in 1960, 5 firms out of 600 controlled 81% of the production. The left-wing repeatedly proposed since 1946-1947 the nationalization of the electric industry, but this was always rejected by the Christian Democratic party, under the pressure of the power monopolies.[9] When this confrontation came to an end in 1948 with the defeat of the left, the minister of industry, Ivan Matteo Lombardo, authorized the escalation of electricity rates up to 2,300%, along with a series of other concessions. Lombardo was a member of the Social Democratic Party and a follower of Saragat, whose crucial role in connection with American interests we will deal with later on. Thus every initiative in the nuclear field – from the nuclear law to the study of reactors – was seen with suspicion: even the construction of nuclear plants was opposed by power monopolies, at least until they realized (around the mid-1950s) that public companies (CNRN, ENI) would start building them. Then they also rushed in, with the intent of influencing the by then urgent nuclear law.

There were however in post-war Italy other economic and political forces that supported different alternatives for the development of the country, aimed at achieving a more autonomous and prominent role in the international context. During the 1950s, Olivetti reached a world-leading position in electronic computers. A very relevant initiative was the foundation in 1953 of a State owned oil company, ENI (*Ente Nazionale Idrocarburi*), having the monopoly for the exploitation of national resources, thanks to the media campaign financed by Enrico Mattei in order to promote the discovery of natural gas and oil fields in Italy. Mattei was an innovative and resourceful, though fairly unscrupulous, Christian Democratic manager,[10] who opposed the schemes of private industry. He also developed a direct, deeply innovative policy towards oil producing countries (the so-called "fifty-fifty" contracts), opposing the oligopolistic international policy of the "Seven Sisters" which dominated the world oil market.

One must also recall that Italy – a unique case among capitalistic countries – had inherited from fascism also an important economic state sector,[11] with IRI (*Istituto per la Ricostruzione Industriale*, Institute for Industrial Reconstruction), founded in 1933 in order to sustain industrial firms and banks facing economic difficulties.

## 5. Early developments concerning nuclear power in Italy, 1945-1957

### *5.1. From the early proposals to CISE, 1945-1952*

The first ideas about peaceful exploitation of nuclear energy were aired as early as August 1945, in the aftermath of the news about the atomic bomb, by three young Italian scientists: the physicists Giorgio Salvini and Carlo Salvetti, of the University of Milan, and the engineer Mario Silvestri, who had just been hired by the electrical firm Edison and employed in the Technical Department. The Managing Director, Vittorio De Biasi, asked the Department to immediately collate information on this new form of energy.

Around the end of the year Amaldi prepared a report entitled *La Fisica in Italia* (Physics in Italy), in which he stated his opinion on what was needed not only for the future recovery and development of fundamental research in Italy, but also for "the right and proper development of peaceful applications of nuclear physics".[12]

Salvini, Salvetti and Silvestri aroused the interest of professor Giuseppe Bolla, of the University of Milan, and at the beginning of 1946 they drafted a three-stage plan for assembling a research group, setting off a nuclear chain reaction, and building in perspective an experimental nuclear reactor. Bolla visited Amaldi in Rome, who agreed to collaborate, with his colleagues Gilberto Bernardini and Bruno Ferretti[13]. The largest industrial groups in northern Italy were contacted, and the idea took shape of setting out an *ad hoc* company.

These initiatives however had to be kept under wraps since Italy had yet to sign the peace treaty, and was forbidden from embarking in nuclear research. The Italian politicians were absolutely unaware of this kind of problems. Thus, in September 1946 the promoters went to Paris in order to contact the Italian delegation to the negotiations for the Peace treaty and warn them of the risk that the treaty could permanently forbid Italy from working with peaceful applications of nuclear physics.

When it was clear that this danger had been escaped, just before the Peace treaty was signed in December 1946, the limited liability, non-profit company CISE (*Centro Informazioni Studi Esperienze*, Centre for Information, Studies and Experiences) was established in Milan on 19 November 1946 in order to build a nuclear reactor for power generation purposes. CISE was housed in premises of Edison, and the shareholders were some major Italian industrial groups: the electric companies Edison and SADE, FIAT, Cogne (foundry), Montecatini (not long afterwards), Falck, Pirelli, Olivetti (in 1949), Terni (in 1950). The Scientific Committee of CISE, presided over by Bolla, was composed by technicians appointed by the shareholders, and integrated by the three young promoters, and the physicists Amaldi, Bernardini, Ferretti, and Giovanni Polvani (University of Milan). Since it was evident to everybody that the birth of a domestic electro-nuclear industry was too big an endeavour for private companies, and the intervention of government was necessary (although feared), the president of the National Research Council (CNR), Gustavo Colonnetti, was included from the outset as an observer.

For more than five years CISE was the only body in Italy which dealt with research on the peaceful applications of nuclear energy. In the early years CISE had to devote itself mainly to the training of qualified personnel. By the end of 1951 CISE (despite a transitory crisis due to the hesitations of some partners) had built a pilot plant to make heavy water through electrolysis, created an experimental uranium metallurgy plant, undertaken in its laboratories important measures on uranium fission, developed leading-edge electronic instruments, besides training specialized personnel.

In the meantime, between 1947 and 1951 the second of Amaldi's aims was fulfilled, with the creation of centres of CNR for fundamental nuclear research in the Universities of Rome, Padua, Turin and Milan, and the final establishment in 1951 of the National Institute of Nuclear Physics (INFN), also pertaining to CNR.[14] The presence of Amaldi in CISE avoided interferences and overlapping among programmes of fundamental and applied nuclear research: in fact, a division of duties was established between fundamental research, centred in Rome, and applied research in the North.

The access of Italy to NATO in 1949 paved the way to collaboration with the United States in peaceful nuclear technology: this was made effective in 1955 with a bilateral agreement, and came into force in July 1956. Previously, the main supply of uranium, around 300 kg, had arrived from Spain in 1948.

Despite this, repeated attempts at making aware the Italian government of the importance of peaceful nuclear technology were unsuccessful. In 1950 for first time CISE unsuccessfully attempted to involve directly the Ministry of Defence and the army, raising the criticism of Amaldi, who was worried that the initiative could pass into the hands of the military.[15] On the other hand, the Italian military showed their ridiculous backwardness in nuclear matters with the alleged experiment of an "H bomb" carried out in the shooting range of Nettuno, near Rome, in 1952, when neither the Americans nor the Soviets had yet produced it![16]

## 5.2. The first State initiative, CNRN, 1952-1955

At last, a receptive political interlocutor was found in the Minister of Public Works, Pietro Campilli: in 1952 the critical economic situation of CISE was rebalanced, and on 26 June 1952 a National Committee for Nuclear Research (CNRN) was established, parallel to similar institutions in other countries.[17] But, in order to bypass thorny discussions in Parliament and opposition inside CISE against any kind of State initiative CNRN was not established with a law, but by a Prime Minister's Decree. This solution deprived the Committee of legal status, an aspect that was to generate several problems in the future. CNRN received state financing from IRI and the Ministry of Industry through CNR, though it was not an organ of CNR. It financed and coordinated the activities of both CISE and INFN. It was provided with a budget of one billion Italian Lire, higher than the whole budget of CNR (while the budgets of the American AEC and the British AEA were equivalent to about 800 and 110 billion Lire respectively). Nevertheless, the establishment of CNRN was considered a great step forward. It was chaired by Francesco Giordani, a chemist who had been president of IRI and CNR during the war, with a southern mentality and hostile to the northern private industry. CNRN included, among others, Amaldi (who left CISE), Ferretti, De Biasi, Arnaldo Angelini, vice president of Finelettrica, and a young geologist, Felice Ippolito, who, as the youngest, was appointed secretary.

CNRN was since the beginning a place of political confrontation and share-out. Moreover, with all its high ambitions, its legal indefiniteness made it an inefficient administrative entity: the Committee overcame these difficulties through expedients designed to circumvent the legal deficiencies, but these were to emerge later on.

CNRN was a committee with no technical staff, while CISE was a research structure. Nevertheless, the collaboration of CNRN with CISE was marked since the beginning by disagreements and suspicions. There was a clash of mentalities between Giordani and De Biasi. The public companies involved in CISE did not meet Giordani's expectations.

In 1952 CISE submitted to CNRN a proposal for the construction of an Italian 1.000 kW natural uranium, heavy-water reactor,[18] but it was not taken into consideration. Only later Giordani made the request for a higher power reactor, and CISE began the work for the already mentioned project Cirene (*CIse.REattore.aNEbbia*, CISE fog reactor). Moreover, Giordani initially opposed CISE's request for financing, objecting that it was a private institution, and when this was resolved the funds were in any case insufficient. In the meantime, other private nuclear companies were established, backed by powerful industrial concerns. FIAT and Montecatini jointly created the *Società Ricerca Impianti Nucleari* (SORIN), and the electric companies the *Società Elettronucleare Italiana* (SELNI). On the other hand, the construction of a major public research centre began at Ispra in 1955.

A major impulse in nuclear matters was no doubt provided by the launch of the "Atoms for Peace" campaign in 1953. Even in Italy decisions took shape, despite the unsolved contradictions.

## 5.3. Missions in the United States and purchase of reactors, 1955

In March 1955, CNRN sent a delegation to the US with the aim of purchasing a supply of heavy water, of negotiating the purchase of a 1.000 kW, CP-5 *research* reactor fuelled with 20% enriched uranium, and of exploring the possibility of the supply of a *power* nuclear plant.[19]

This was indeed a radical change of objective compared to the previous programme of CISE, which focused on the project of a domestic reactor. This turned instead to a dependence on foreign technology, and was considered a betrayal by Edison, worried that the public sector could develop on its own the production of electronuclear energy.

This initiative prompted an immediate reaction by the electrical industry, which in October 1955 sent to the US another mission, composed by Giorgio Valerio, General Manager of Edison, Franco

Castelli and Silvestri. This mission also aimed to discuss the purchase of a nuclear reactor, after some contacts had already started in 1954: such initiative raised some surprise even inside the AEC, besides the opposition from CNRN, which urged a legislative initiative from the government.

At the Geneva Conference on nuclear power Italy presented 6 communications (4 from CISE).

In July 1956 CNRN president Giordani resigned, trying to force the Government to overcome the situation of uncertainty and financial difficulties of the Committee (he was appointed president of CNR). The Secretary, Felice Ippolito, took the position of General Secretary. Prime Minister Segni was inclined to dissolve CNRN, but Ippolito succeeded in strengthening it. CNRN was officially renewed with a Prime Minister's decree in August 1956. Since the new President, a trustworthy Christian Democratic senator named Basilio Focaccia, was completely unread in nuclear energy, Ippolito had more freedom in governing the Committee and pursuing his own views: a new phase began, fraught with crucial events. With the renewal of CNRN, De Biasi was excluded.

In this period the CP-5 research reactor was finally purchased by CNRN from the US (though some, like Edoardo Amaldi and Mario Ageno, pushed for a British choice): it was very expensive ($3 million) and became operative in 1959 at the new centre in Ispra, with a power of 5.000 kW and the name Ispra-1. It showed immediately heavy design shortcomings.[20]

The construction of the mentioned research centre at Ispra raised immediate problems for CNRN. Lacking legal status, it had to create an ad hoc joint-stock company in order to acquire the ground. CISE was involved in the construction, but the tensions with CNRN flared, leading in September 1957 to a break in all relations between the two organizations. CNRN decided to continue the project alone. But again CNRN had to create a joint-stock company (NUCLIT), which took on a large number of the staff previously employed by CISE, which consequently had to recast its programmes and activities. CNRN finally took a dominant position in Italy's civil nuclear industry, assuming also technical tasks. The centre of Ispra was finished in 1959, but it was almost immediately made over to Euratom, together with the CP-5 reactor, for a Community centre. It was probably a far-sighted choice on Ippolito's part, judging that the burden was too high for CNRN.

Contradictions emerged also at a European level. In 1956 a commission composed by Louis Armand, Franz Atzel and Giordani, president of CNRN, proposed a gigantic programme for installing by 1967 a total nuclear power capacity covering ¼ of the electrical output of the six States, purchasing reactors from the US. On the contrary, in 1957 CEE established Euratom with the opposite aim of developing an autonomous and competitive nuclear capacity. The same 1957 was the year of the secret initiative of France, Germany and Italy with the ambitious purpose of developing European nuclear armaments,[21] which was abandoned when De Gaulle decided to develop a French *force de frappe.*

Also the Italian military tried again to enter the nuclear programmes. An attempt to condition the national choices and to join CNRN in 1954 was thwarted by Giordani. In 1956 CAMEN (Centre for Military Applications of Nuclear Energy) was established near Pisa,[22] with the mission of training personnel and studying nuclear propulsion and nuclear armaments. The military planned the purchase of a nuclear reactor in the US through CNRN, clashing with Ippolito. Through the Ministry of Defence and collaboration with the American embassy in Rome, in 1957 they purchased a pool-type reactor,[23] an untenable choice for military applications! The reactor was shut down in 1980, with subsequent controversies on the (classified) decommissioning and delivery of radioactive wastes. CAMEN later changed its name into CISAM (Inter-service Centre for Military Studies and Applications): in 1973-1975 classified tests were performed of a nuclear capable missile. Italy was to join NPT in 1975.

## 6. The three nuclear plants, 1957-1964

After these preliminaries, among these deep contradictions, and without any general strategy, between 1956 and 1958 the contracts were signed for the purchase of three nuclear power plants, by

Edison, CNRN and ENI in close sequence. They were in fact three uncoordinated, completely different, and competing designs, but they were to project Italy in 1964 to the third place for electro-nuclear production in the world, behind the US and UK: a sudden emergence that was the contingent, almost paradoxical, result of the deep contradictions, even of the backwardness of the country. Around 1957 there were barely a few prototypes of power nuclear plants working in the world.[24] The choice between different models was objectively very complex, and the available experience very restricted. The choice also had heavy political implications: for instance, the choice of an enriched uranium reactor would have determined a strict dependence from the US for the fuel, but a different choice would clash with American interests. The lack of a national legislation, hindered by private industries, created big problems (localization, licence, control, protection, management of the cue of the nuclear cycle), and even obstacles in the bargaining with the United States.

What's more, in all three cases the Italian purchasers ordered plants that were still in an experimental stage, for which no prototypes were actually up and running, and which were built concurrently with the first reactors of their type in the US and UK. The technology was immature, and consequences were to become apparent during plant operation.

We have seen how Edison's visit to the US in 1955 was a reaction to the previous mission from CNRN. When Edison was the first to order a light-water nuclear power plant from Westinghose, this was an attempt to prevent the advent of state-controlled nuclear power from paving the way to nationalization of the electrical industry. Without this fear, private industry would not have invested any money in nuclear energy. In 1957-58 Edison, through SELNI, applied to the Italian Government for a guarantee on exchange rates, in order to protect the loan from future rate fluctuations.[25] In winter 1957-58 the loan had been formalized with the Export-Import Bank for $34 million: however Ippolito pressed the Ministry of Industry against the lending, which in fact was denied. De Biasi asked in vain to sack Ippolito. The construction of the Edison nuclear plant was delayed, and it was preceded by those built by the state groups IRI and ENI. The contract was signed only in 1957 with Westinghouse, for a 242 MW reactor, whose prototype was the Yankee one at Rowe, Massachusetts. Only in 1957 CNRN approved the project, subject to a safety report by the company, and only in 1960 the location was decided, at Trino Vercellese, near Turin: a previous attempt to locate the plant at Moneglia, near Genoa, was thwarted by the first instance of popular opposition. The plant reached criticality in June 1965, was first connected to the network in October 1965, and began commercial generation in December 1965. The final cost of the plant was of $41 million (against the $34 million loan).

In October 1956 both IRI and ENI expressed the intention of building nuclear plants, in Southern Italy (as national entities). In 1957 the International Bank for Reconstruction and Development (IBRD, of the UN), together with the *Banca d'Italia*, proposed to the Italian government a common study for a nuclear plant. The agreement was signed in July 1957, and the construction (project by ENSI, *Energia Nucleare Sud Italia*) was assigned to Finelettrica (IRI) through the public company *Società ElettroNucleare Nazionale* (SENN, drawn out of SELNI). Nine companies participated in the international tender (four American, four British and one French). In September 1958 an American reactor was finally purchased, a 150 MW, BWR light water reactor from General Electric. It was built at Garigliano, Southern Italy, with funds from the *Cassa per il Mezzogiorno* (Fund for Southern Italy) and a financial support of $40 million from the World Bank.[26] The plant reached criticality in June 1963, was initially connected to the grid in January 1964, and began commercial generation in May 1964 (Figure 2): it was to be shut down for good in 1978 due to an accident.

The third project deserves specific attention. It was built by ENI, whose president, Enrico Mattei, had created in 1956 a specific section, AGIP-Nucleare: for the construction, the SIMEA company was created (*Società Italiana Meridionale per l'Energia Atomica*). Mattei's choice is remarkable since, in accordance to his strategy of gaining autonomy from the United States, he chose a British gas-cooled "Magnox" reactor, fuelled with natural uranium. The plant was built in Southern Italy, near Latina. Its total cost was around $60 million, and in fact it was the first one to enter into function: criticality was reached on 27 December 1962, electric generation began in May 1963, and commercial production in January 1964. At that date it was the most powerful nuclear reactor in Europe.

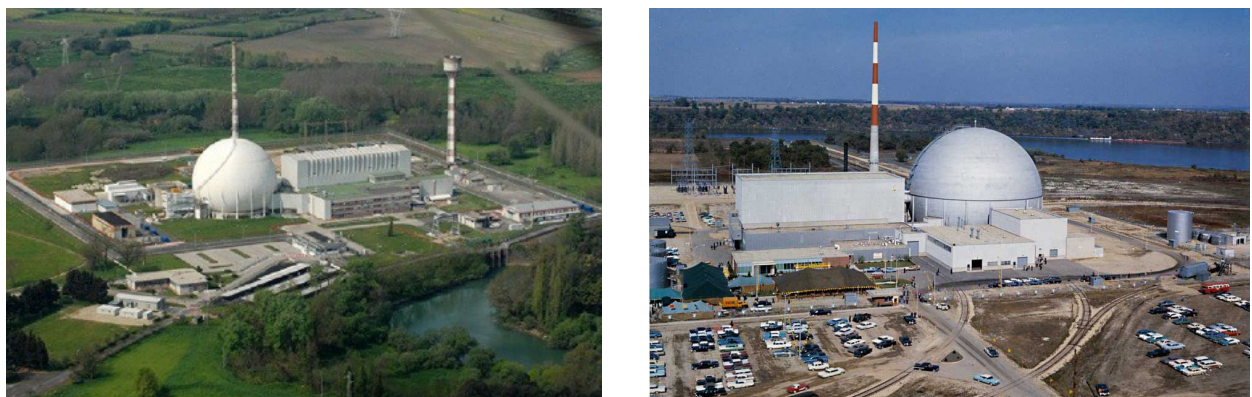

**Figure 2**. The Garigliano NPP – and his American twin, Dresden1 – were characterized by a containment sphere to prevent radioactivity release. In reality the containment sphere does not ensure a perfect seal, so this solution was abandoned after these two prototypes, and was replaced by more efficient containment structures.

The profitability of these projects, and the cost of the electric power produced, seem far from clear, and caused heated and crucial debates (Section 9). A "partisan" diagnosis, from a CISE representative, maintains that: «According to official estimates the cost of the electric energy Kwh was Lire 7,80 (Latina), Lire 7,20 (Garigliano), Lire 5,40 (Turin), compared with a cost of traditional energy lower than 5 Lire. This means that the annual burden for Italy is around 7 - 8 billion Lire».[27]

## 7. 1960, the ambiguous transformation of CNRN into CNEN and the proliferation of nuclear centres and projects

A CNRN "white paper" of 1957 was practically a five-year plan for nuclear research. In subsequent years CNRN got to administer significant amounts of money and employed around 1700 people. Its activities encompassed a wide range of ambitious initiatives and projects, from the development of reactors and the fuel cycle, to the training of scientific and technical staff, the diffusion of information and scientific culture, and fundamental physics. The latter field absorbed around 20%, through the INFN, whose main commitment was the construction of a large electron synchrotron at Frascati laboratory, near Rome.

The legal regulation of the field at a national level remained however far from clear. The opposition against an organic nuclear act caused the failure of four bills during the 1950s. The European (Euratom) and international (IAEA) developments after 1957, as well as the gift of Ispra to Euratom by Ippolito in 1959, reinforced support for national regulation, in spite of the resistance from the private electrical industry. Moreover, in 1959 CNRN was under extension and faced again financial difficulties. The new Ministrer of Industry, Emilio Colombo, submitted another bill, which met again strong opposition. But further postponement was infeasible, and a transitional law was approved in August 1960, establishing the National Committee for Nuclear Energy (CNEN, *Comitato Nazionale per l'Energia Nucleare*).[28] It was not a merely new name for CNRN, since CNEN was assigned new tasks, such as the technical control and testing of all the nuclear plants under construction, and it was granted a legal status, overcoming the anomalies of CNRN: nonetheless, not even CNEN was a public economic corporation enabled to carry out entrepreneurial initiatives. CNEN was chaired by the Minister of Industry (who however assumed the double role of inspector and inspected), and was run by an Executive Committee. But the five-year plan for nuclear research underwent further delays, creating recurring financial problems to CNEN. Moreover INFN remained in a state of legal uncertainty, and of tension with CNEN.

Obviously CISE expressed a very negative opinion on this act, since CNEN left it no elbow room whatsoever. The more so, since Ippolito was appointed Secretary General of CNEN. As a matter of fact, Ippolito was given a free hand in the administration of CNEN. This started a phase of great impulse and dynamism in the management of the Committee and of the whole Italian nuclear programme, a sort of (however brief) "age of Ippolito". There is no doubt that in the confused institutional status of CNEN, and with the chronic stickiness of Italian bureaucracy hindering scientific research, Ippolito's administration adopted unscrupulous procedures, at the limit of legitimacy, in order to circumvent bureaucratic complications and speed up the decisions. But neither is there any doubt that this style was approved and admired in the milieu of research, and the "age of Ippolito" was to be recalled as a model of efficiency. It is important to remark that even the political class in the government was informed, or aware, of possible administrative irregularities in the management of CNEN, but did not intervene: the storm was to explode in a sudden and unforeseen way three years later (Section 9).

Actually, the new tasks of CNEN multiplied the ambitious nuclear programmes and centres, which were only in part already provided by CNRN: Italy, although still lacking an organic organization and legislation, apparently acted as one of the most ambitious actors in this field! In fact, the overall programme aimed at reaching a complete autonomy for the Italian nuclear industry, from the design and construction of reactors, up to fuel fabrication and reprocessing, i.e. the full fuel cycle. However, as we have remarked, such an ambitious programme was not an exception, but fitted into a project pursued by a part of the Italian leading class, which meant to make the country a leader on the world market, both economically and technologically. In particular, not only the goal of energetic independence, but also the unscrupulous approach was common to both Ippolito and Enrico Mattei, although the latter was rather a business leader.

Actually, the *first five-year plan* (1959-1964) for CNEN, financed with 80 billion Lire for the first year, planned the design and testing of four types of reactors: boiling water, cooled with organic substances (PRO, *Progetto Reattore Organico*, with a cooling mixture of biphenyles and triphenyls), liquid metal, and high temperature gas. The PRO was a joint project between Agip Nucleare (ENI Group), Fiat and Montecatini, through AGIP-Nucleare and SORIN, including a partnership with American companies (Atomic International, Martin Marietta), coordinated by CNEN via research contracts. The project was abandoned during the 1960s (see below). One must remark that some of these projects are not yet satisfactorily working even today in the most advanced nuclear countries!

On the side of fuel reprocessing, Italy was already taking part in the Eurochemic experimental plant near Mol, Belgium (an OECD joint venture), through CNEN and SORIN, while the EUREX (Enriched URanium EXtraction) pilot plant was under construction at the Research Centre of Saluggia, in northwest Italy:[29] in the forthcoming years changes in Eurochemic ended up making the Eurex plant obsolete.[30] Moreover, the centre of Casaccia had been established near Rome, and provided with some research reactors.

A programme was also put in place for the uranium-thorium cycle (PCUT, *Progetto Ciclo Uranio Torio*)[31], in collaboration with the American AEC, aiming to design, build, and run both a chemical separation plant and a pilot integral-cycle plant: two centres were built, at Trisaia in the South and, some years after, near the Lake Brasimone, on the Tuscan-Emilian Apennines. The Trisaia plant (ITREC) stood for many years as the only one in the world designed to recuperate U-233 from Thorium: in those times it was an original design, but like the PRO, the PCUT was to be abandoned during the 1960s. After the shutdown of the Italian nuclear programmes (Section 9) Trisaia became a "nuclear rubbish bin", when 84 fuel elements from the Elk River American nuclear plant were purchased in 1967, while the U-Th cycle was abandoned worldwide, with the exception of India. Sixty-four fuel elements remained in the site, with no more possibility of reprocessing, a paradoxical inheritance.

However, all these centres and programmes were to leave a heavy inheritance when the whole Italian nuclear programme came to a stop!

### 8. 1962, the nationalization of the electric industry (ENEL) and the Nuclear law

One of the indisputable merits of CNEN's approach to the nuclear programmes was that it involved collaborations with several industrial sectors, giving them a stimulus, and contributing to divide the industrial front opposed to the Nuclear Law. On the other hand, both the development of the nuclear programme and the developments in Europe posed the increasing need of overcoming the fragmentation of the Italian electrical system. This concerned both domestic production and the unification of the national electric grid on the one hand, and, on the other hand its connection to the European grid. An important step was the unification of electric fares in May 1961, which also strengthened the need of extending the grid to all users, by building an Apennine ridge power line, and providing for a national power reserve to meet the peaks in demand. Which private producer could assume the burden of building plants to be up only in these cases?

The definition of a Nuclear Law became therefore strictly intertwined with the harshly debated problem of the nationalization of the electric industry.[32] This measure was strongly supported by the Socialist party (PSI), and was finally approved by the first centre-left government in December 1962, with the creation of ENEL (*Ente Nazionale per l'Energia Elettrica*). It was really an epochal step in Italian history. The Christian Democratic Party, inside which fierce power struggles were underway, accepted nationalization out of calculation, or at a Hobson's choice. On the other hand, the electric utilities were indemnified with pharaonic amounts of money. The underlying idea was that the firms would use such huge amounts to develop other sectors, and give a tremendous impulse to the Italian economic system: unfortunately, the Italian managers proved frightfully incapable, and squandered such a richness in the following years.[33] Ippolito himself correctly remarked that the state could have bought the electrical utilities at a very low cost after the war, whereas it supported their recuperation: in that case Italian history could have taken a completely different turn.[34] In 1962, moreover, the electrical industry had already exploited the main benefits from the existing resources, so that nationalization resulted in a big business.

The first centre-left majority finally approved in December 1962 the nuclear bill, already submitted since January 1961, which established that the production of nuclear energy was a prerogative of the state, or of state-controlled stock companies. The *second five-year plan* (1965-1970) for CNEN had already been approved in October 1962. It confirmed all the previous programmes, and provided for the installation by 1970 of 1.000-1.500 MW with the design and construction of 2-4 nuclear plants. The division of duties between the nuclear body and the electricity organization was established. CNEN was assigned – alongside its duties of research, design and development of reactors – the responsibility for monitoring plant safety through its Safety and Protection Management Office (DISP, *Direzione Sicurezza e Protezione*), which, as a matter of fact, has made up for the absence in Italy of a Nuclear Safety Agency. ENEL was put in charge of developing nuclear power within the framework of the national electricity system, with decision-making capacity on all aspects of power stations construction. The three nuclear power stations were handed over to ENEL.

At the beginnings of 1963 Ippolito was appointed to the board of directors of ENEL, though the law which had established CNEN did not allow this. This fact was to have important implications in the forthcoming events.

### 9. The "Ippolito trial" (and other disquieting events), 1962-64, and the halt to the Italian ambitions of autonomous development

Behind these apparently successful events, however, the political and economic circles, both national and international, were very concerned about Italian development policies, and perceived a great danger for their privileges and business. Starting exactly from 1962 a succession of disquieting events, which could hardly be considered accidental, systematically eliminated all the most resourceful leading players on the Italian scene and stopped the most advanced experiences, not only the ambitious nuclear programmes. *One can hardly avoid the impression that the history of Italy came to a crossroads, and that the most conservative forces (among which mafia) and the most powerful international interests succeeded in eliminating every chance of progressive development, growth and competitiveness in Italy*: in the subsequent decades a lot of "mysteries" and unsolved assassinations, massacres and attempted *coups d'état* have stained with blood the history of the country. In 1974 Pier Paolo Pasolini warned: "I know … but I have not the evidence".[35] Since the spring of 1972 Pasolini was working on a novel, *Petrolio* (Oil), which was to be a pitiless and substantiated accusation about «the crisis of our republic, with oil in the background as the big protagonist … our sufferings, our immaturities, our weaknesses, and the state of subjection of our bourgeoisie, of our presumptuous neocapitalism» (from a statement of the writer Paolo Volponi)[36]. When Pasolini was murdered, in 1975, the novel was still only a draft, which was published in 1992 with a white cover:[37] there was moreover the unsolved mystery of the manuscript of one chapter, which disappeared!

Already in 1960 there had been an attempt at a fascist revival with the Tambroni cabinet and subsequent bloody street reactions. It has been proved that among the political and financial supporters of the Tambroni cabinet there was a part of the private electrical industry, determined to stop at any cost the discussion that had started in Parliament on the nationalization of the electric industry: there was talk of slush fund, and Giorgio Valerio's involvement.

On October 27, 1962, Enrico Mattei – the unscrupulous and controversial oil bargainer with the Arab countries, who was circumventing the "Seven Sisters" – died in a crash of his personal aircraft. The cause has never been clarified, but was almost certainly an attempt on his life.[38]

Just a couple of months before, on August 11, in the midst of vacations, the influential Italian newspaper *Corriere della Sera* had published a note by the already mentioned President of the Socialist Democratic Party (PSDI), Giuseppe Saragat, with the title: "Electricity and nuclear energy: squanderings denounced by Saragat": this ambiguous individual, strongly linked with the American circles, affirmed that ENEL was a good company, while CNEN's General Secretary, Ippolito, administered in a quite questionable way the funds from the State; and running the nuclear plants was absolutely uneconomic in comparison with traditional plants. This preliminary attack was therefore against nuclear energy, and Ippolito, who was its most combative supporter.

It is impossible, and out of place here, to enter into the details of this affair, which has still many obscure aspects, basically related to power struggles, mainly inside the majority Christian Democratic party.[39] Its substance however is very clear: *the final aim was to stop, through the person of Ippolito, the Italian nuclear programmes and projects!* In Italy bureaucracy has always been extremely complicated, and difficult to respect in any detail: in any case, no matter how illicit Ippolito's administration may have been, this could not have justified the stop to the nuclear programs. For such irregularities, the majority of the Italian ruling class should be in jail. However, this is precisely what happened!

There was a committee of inquiry, and a shadowy penal case against Ippolito, who was sentenced to 11 years' custody (unanimously deemed an exaggeration). He was the sacrificial lamb, while the politicians who formally had to direct CNEN and authorize Ippolito's expenditures were not even mentioned. The Italian scientific community was in its majority sympathetic with Ippolito: Edoardo Amaldi, the most influential scientist, publicly attacked Saragat; a distinguished scientist as the geneticist Adriano Buzzati-Traverso, on the weekly magazine *L'Espresso, spoke* of "a new witch-hunt ongoing in Italy". But all that was irrelevant: the problem was a political one.

Subsequently Saragat was elected President of the Republic. In later years Ippolito was rehabilitated, was granted pardon by Saragat, and even bestowed "for his scientific merits" with the Cross "for merits" of the Republic. But in the meantime the Italian nuclear projects had undergone a sudden stop.

Actually, the Ippolito and Mattei cases were not isolated. In 1962 Domenico Marotta, although already retired, was indicted for administrative irregularities:[40] he had been an eminent chemist and manager, and as Director of the *Istituto Superiore di Sanità* had brought the Institute to reach a high international standing, with two Nobel Laureates working for it, Ernst Boris Chain and Daniel Bovet.

One more very meaningful case was that of the Olivetti company, which had reached a world leading position in electronic computers, pioneering the personal computer with the model "Olivetti-101": in fact, in 1964 the control group of the firm – composed by Fiat, Pirelli and two public banks – decided to transfer, in the total indifference of Government, the Electronic Division to General Electric! Even here there had been in 1961 a "death" in a car "accident", that of the outstanding engineer Mario Tchou, director of the Olivetti laboratory.

As a matter of fact, these events marked the beginning of a sharp decline in the Italian post-war aspirations to an advanced scientific and technical development, which was barely taking shape among deep contradictions. Throughout the 1960s Italy got the green light to advanced research in the field of high-energy physics and accelerators, but was long cut off from more applied fields, such as solid-state physics and nuclear reactors.[41]

## 10. The fourth nuclear plant

The developments of the inquiry on Ippolito and his trial produced a deep bewilderment in the world of Italian scientific research, and a strong dejection among CNEN's staff, who felt under indictment. Ippolito represented efficiency, struggle against bureaucracy, and had become a centrepiece of the whole system of Italian research, which unanimously sided with him. The October 1963 report of the inquiry committee of the Ministry was perceived as an attack against Italian technical-scientific culture. Anyway it marked the final end of Italian nuclear ambitions.[42]

An immediate consequence was a stiffening of the administrative procedures, and a slowdown of CNEN's activities. The new Moro government (December 1963) speeded up the funding to CNEN. But a report issued by Minister of Industry Giuseppe Medici on the nuclear issues (June 1964: Medici 1964) concluded that the two "pilot experiences" (PRO and PCUT) had not kept their promises (but, as we have remarked, all five failed). A technical committee led by Mario Silvestri (from CISE) curtailed CNEN's activities, closing a number of programmes that Silvestri had opposed in previous years. In the new CNEN steering committee, appointed in 1965, Carlo Salvetti was designated as Deputy Chairman, and Angelini was the sole member confirmed from the previous committee. Nevertheless electrical consumption in the country was growing, and in 1966 both Salvetti and Angelini advocated a renewed nuclear power station building programme.

At any rate, ENEL acquired the property of the nuclear plants, absorbing a large part of their technical staff. The banner of nuclear energy was taken up by the Italian Communist Party (PCI) and the unions: the PCI engaged Ippolito, but it had not clear ideas on nuclear matters, like the majority of the political class.

In any case the second five-year plan had become outdated, since nuclear technology had deeply evolved since the prototype reactors.

In 1967 ENEL decided to build a fourth Italian nuclear plant, defined the technical specifications, and called for tenders, receiving several bids (PWR, BWR, Gas, and Candu reactors), from American, British, and other European utilities. There were political pressures and compromises. The "American party" prevailed against a European choice. Finally in 1969 an 840 MW, BWR reactor was chosen, from General Electric and *Ansaldo Meccanico Nucleare*: the reactor actually was a somewhat hybrid, transition model between nuclear power plant generation I and II. The PCI was left with the "swindle" of Cirene as a sop.

Meanwhile, ENEL began to plan a fifth power plant, setting in 1968 the procedures for its construction. However, on the one hand in 1969 safety regulations for new nuclear plant construction and operation were tightened in the US, with substantial growth of costs. On the other hand, ENEL had to cope with growing financial difficulties, since it had inherited the burden of paying compensation to the private electric companies, without receiving adequate financing: the Public Accounting Office expressed the opinion that its debt should not be allowed to grow any further. Funds were insufficient to order nuclear power plants.

The construction of the fourth plant began at Caorso (Northern Italy) in 1970, by a consortium formed by ENEL-*Ansaldo Nucleare*-GETSCO (General Electric Technical Services Company). Works were to be completed by 1975, but they suffered delays and cost increase: Caorso's first connection to the grid took place in 1978 and commercial operation began in 1981.

In 1971 a rearrangement of CNEN was approved by law, but its implementation was delayed. In December 1971 procedures began for ordering a fifth power plant with a power capacity between 800 and 1.000 MW, and a call for tenders started in December 1972.

### 11. The 1973 oil crisis and the "oil scandal". The troubled path of the Italian nuclear projects, the  National Energy programmes (PEN), and the growth of popular and environmentalist opposition, until the Three Mile Island accident

The 1973 oil crisis subverted all the concepts and world forecasts on energy resources, production, and consumption.

In Italy, at the beginning of 1974 some magistrates investigating the hoarding of oil during the Kippur war, found eloquent oil industry documents, implicating all the Italian political parties (except PCI), ENEL's managers, ministers, in illegal procedures and subsidies in favour of the oil industry.[43] As it often happens in Italy, in spite of corruption for billions of Italian Lire (millions US \$), in the end there was no legal consequence for politicians and managers. The contrast with the sternness in the Ippolito affair could hardly be more striking!

The scandal moved those who had stopped nuclear energy to re-emerge. In December 1973 ENEL, upon assurances from the government on the adoption of a new legislation, decided to buy not one but two power plants, and in summer 1974 orders were placed for four. In August 1975 the government passed a law that regulated localization procedures, in relation to the recently published American "Reactor Safety Study" (known as "Rasmussen Report"), which set a safety zone of 16 km radius around nuclear plants. On the basis of this law, ENEL proposed to place the four plants in Central Italy, two in Lazio (localized later at Montalto di Castro, see below) and two in Molise.[44]

After oil prices suffered a further hike, CNEN submitted a National Energy Plan (PEN) that was approved by CIPE (*Comitato Interministeriale per la Programmazione Economica*) in December 1975. The Plan foresaw different future scenarios for energy demand in Italy for electric energy, it envisaged the installation in the period 1983-85 of a nuclear power capacity of 13.000-19.000 MW (up to 20 power stations), and further plants for a total capacity between 46.100 and 62.100 MW by 1990.

In 1976 the study of environmental impact was presented for the localization of 2.000 MW nuclear power plant (BWR type) at Montalto di Castro (Maremma).

In the meantime, a new factor was radically changing the situation with respect to the previous decade, since the Italian population became increasingly conscious and sensitive towards the problems of health and the environment, as well as the dangers of nuclear energy. In the first respect, the worker's struggles in defence of health in the factories, which started during the "Hot Autumn" of 1969, extended to the population, anticipating the environmentalist movement. In the second respect, in the late 1970s the first big nuclear accidents began to shake the previous absolute (and largely uncritical) confidence in the safety of this technology, and to increase the awareness of the extreme seriousness of its potential consequences. As a matter of fact, strong opposition towards nuclear energy began to grow among the populations concerned with possible localizations of nuclear sites, feeding the birth of popular committees, environmentalist associations, as well as opposition from some minority political forces, and even local administrations. The role of "popular experts" began to emerge, and they were involved in animated public debates with technicians of the electric industry or nuclear experts.[45]

Big demonstrations took place at Montalto di Castro, Viadana, Suzzara, San Benedetto Po (in Lombardy, when the localization of nuclear plants was proposed there). The associations WWF and "Italia Nostra" collaborated, produced documents, and summoned meetings and debates;[46] the Lombardy Region appointed a study commission on nuclear plants, and turned to the *Istituto Superiore di Sanità* for advice. In these harsh debates, one should take into account the very peculiar structure of the Italian territory, with many mountains and few flat lands, and a high density of population and of towns.

However, the majority of the political forces and the unions were strongly in favour of nuclear energy, including the majority of the Communist Party and the left-wing Union CGIL. Anyway, in response to these movements the political debate grew: the Industry Committee of the Parliament held a fact-finding inquiry, which confirmed the government's focus on nuclear power (April 1977), and a Parliamentary debate took place.

The third Andreotti cabinet (which included the PCI) called for an immediate start on power stations, and the Ministrer of Industry, Donat-Cattin, issued an ultimatum to the Regions asking them to indicate the sites for the construction of 20 nuclear plants. ENEL sent out calls for submission of technical bids for a further eight 1.000 MW units.

In December 1977 CIPE adopted a revised PEN, providing for the immediate construction of "only" 12-13 nuclear plants, and postponing the remaining 8 until after 1985. The estimates of this plan on future electric energy requirements would later turn out to be exaggerated. It was an "electric" rather than an "energy" plan, and several reported data were wrong, or inconsistent. In response to this, popular protests and marches grew even more. The more so when Prodi, Minister of Industry in the 4th Andreotti cabinet, on 19 February 1979 authorized the construction of the nuclear plant in Montalto di Castro: just before the Three Mile Island accident, on 28 March 1979! In the same days the film "Chinese Syndrome", with Jane Fonda, came out. In the meanwhile, in August 1978 the Garigliano nuclear plant had been shut down for good after several accidents. One must remark that by the late 1970s the construction time of a power plant was increasing, and nuclear industry development was experiencing a general slowdown.

When the subsequent 5th Andreotti cabinet fell, the political elections were preceded by a huge national protest in Rome on 19 May 1979.

One might ask who should have paid for these gigantic projects. The funds were to be anticipated, at a high rate, by the American Export Import Bank, and in his 1977 visit to the US Andreotti got the support of the Monetary Fund, offering both political (no access of PCI to government) and economic (unpopular measures) guarantees. The Italian industrial groups, both private and public, were competing for the different patents (Westinghouse's PWR, General Electric's BWR, Babcock and Wilson's PWR, and Canadian Candu). However, the Italian industry had reached a certain development: a 1978 *Confindustria* (Italian Industrial Federation) document ascertained that in 1977 and early 1978, Italian electromechanics companies had won more than 40% of all international electric power station calls for tenders. By the end of the 1970s Italy's nuclear industry had acquired a lasting configuration, in which ENI (*Fabbricazioni Nucleari*) focused on fuel-related provisioning, and IRI-Finmeccanica (AMN) was responsible for building plants under license from General Electric.

Moreover, since 1973 a challenging and expensive international collaboration began, when ENEL entered with a 30% participation in a joint venture with French EDF and German RWE to build "fast" breeder reactors[47] (NERSA, *Nucléaire Européenne à Neutron Rapid S.A.*). An Italian contribution to this programme was to be the construction, at the Centre of Brasimone, of an experimental reactor for carrying out tests on fuel elements for fast reactors (PEC, *Prova Elementi Combustibile*).[48] In addition, Italy acquired a 25% participation in the French gas-diffusion enrichment plant Eurodif. Moreover in 1976 the silly project was elaborated of a second enrichment plant to be built in Italy, Coredif, fuelled by four nuclear plants of 1.000 MW each! That was four times the total Italian nuclear power capacity, including the not-yet-operating Caorso plant. Fond ambitions were back, in the typical Italian style of improvisations and contradictions. Consider that France was developing its *Force de Frappe*, and had a strong need of plutonium and highly enriched uranium: nothing similar was happening, thankfully, in Italy!

As for the big oil industries, their role was by no means weakened after the 1973 oil crisis: in fact the "Seven Sister" were increasingly investing in the nuclear sector. The crisis itself had been piloted from New York, in order to make nuclear energy and American oil competitive.

### 12. The Eighties: feverish sequence of committees, inquiries, governmental decisions, movements and protests, until the show down.

The sequence of events became increasingly feverish and excited, and the nuclear problem blew up as one of the hottest in the Italian context.

The main bottleneck in the implementation of the Italian nuclear programmes concerned the localization of the power plants, which was not ENEL's responsibility. The choice was complicated by strong local opposition.[49] While the nuclear programme had broad cross-party support, even many local party exponents opposed the localization of a plant in their area.

The situation grew worse after the 1979 Three Mile Island accident. In the US it was proposed to authorize the localization of nuclear plants farther from residential areas, to provide emergency plans and an evacuation radius in case of accidents of 30-40 km. In almost all the Italian territory tens of thousands would have to be evacuated!

In Italy, on the institutional side, in June 1979 the results of a fact-finding special ecological commission from the Senate got a majority of favourable opinions, with the relevant exceptions of WWF, "Italia Nostra", and Prof. Mario Pavan of the University of Pavia. In December the new Ministrer of Industry, Antonio Bisaglia, appointed a Commission on nuclear safety, which approved a document denouncing the deficiency of the Italian safety rules: but the evaluation was considered weak by the three environmentalist representatives, who presented a minority report.

PEN was successively revised in 1980 and 1981, providing for the construction of nuclear plants for at least 6.000 MW (indicating the Regions of Piedmont, Lombardy, Veneto, Tuscany, Campania, Puglia, and Sicily). The plan introduced the "standard plant" concept, dubbed "Unified Nuclear Project" (PUN), based on the Westinghouse pressurized water technology, conflicting with the previous choice of General Electric boiling water technology for the Caorso and Montalto plants. The allocation of responsibilities was to ENEL for commissioning and systems architecture, CNEN (ENEA, see below) as the monitoring authority, AGIP Nucleare for fuel supply, and Italy's private nuclear companies (through a consortium led by AMN as the "main contractor") for the supply of plant systems and components.

This was to be Italy's last serious attempt at nuclear planning. During the years 1981-1983 popular opposition against nuclear power grew further, supported also by decisions of several municipalities. In order to circumvent this opposition, the way out of economic incentives was tried, with a law of 1983, providing financial support to municipalities that accepted nuclear and thermoelectric plants in their territory (besides nuclear, also coal was pushed by the various PENs). It became clear that out of the four power stations ordered under ENEL's nuclear programme, only the Montalto site had a realistic chance of being completed: construction work continued at an exceedingly slow pace. In these same years Italy had to reduce from 25 to 16,5% its participation in the Eurodif enrichment plant, and to undersell a part of the enriched uranium it had already acquired, following the downsizing of its nuclear ambitions.

In the meantime, in 1982 CNEN acquired the new name ENEA, National Authority for nuclear and Alternative Energies (*Ente Nazionale per l'Energia nucleare e le Energie Alternative*), with a few real changes, but a new research section on renewable energies-- although the new 1985 PEN confirmed 12.000 MW of nuclear power.

ENEA expressed its favourable opinion for the suitability of the sites of Viadana and San Benedetto Po, and ENEL begun the geological tests. Anti-nuclear demonstrations, clashes with the police, and arrests followed. Two municipal referendums were held, at Viadana (1984, 91 % votes against) and San Benedetto Po (1985, 89 % against). In 1985 there was a big march in Rome.

It must be remarked that the anti-nuclear movement was reinforced by the "Euromissiles" crisis. The "Doomsday Clock" of the *Bulletin of the Atomic Scientists* was put at barely 3 minutes from Midnight! The debate on "nuclear winter" grew, the film "The Day After" further impressed the public, and there was strong opposition to the deployment of Cruise nuclear missiles in Comiso, Sicily, to balance the planned deployment of SS-20 missiles in the East European States.

## 13. The last acts of the "drama"

We come to the penultimate act, ironically only 36 days before the Chernobyl accident! On 20 March 1986 CIPE approved the 4$^{th}$ PEN (if the progressive number makes sense), providing only for the construction of the 2.000 MW plant at Montalto di Castro, plus 2.000 MW more at Trino Vercellese, in Piedmont (never begun), and the localization by 1986 of two more plants of 2.000 MW each, respectively in Lombardy and Puglia; in addition it provided for the acquisition of 400 MW from the 1.200 MW fast reactor *Superphénix* under construction in France, a very unfortunate project, for which Italian taxpayers have paid out 30 % for two decades (to the French government and military the project provided the management of the plutonium cycle, and the supply of plutonium during the construction of the *Force de Frappe*).

On April 9-13, 1986 the PCI held its XVII$^{th}$ congress, in which an anti-nuclear motion was presented and got almost 50 % of votes.

Barely two weeks later, on 26 April 1986 the Chernobyl accident happened. It raised a deep impression and a huge concern for the behaviour of the "Chernobyl cloud", and the public debate and polemic revived. Local and national demonstrations (Rome, 10 May) proliferated. In July the gathering of signatures for a national referendum began. In October, after a huge demonstration at Montalto di Castro, the Craxi cabinet decided the stop to the building site, and called for a big Conference on Energy, which was held in February 1987, without any important result.

The referendum took place on 8-9 November 1987, and was the prologue to the last act of the Italian nuclear "comedy". The campaign was harshly fought by the movements and their exponents against the large majority of the scientific and technical milieus, which kept up their firm faith in science and technology. Since the Italian law formally allows only for questions concerning the abolition of specific laws or regulations, the referendum formally could not decide the shut-down of the Italian nuclear programmes. The questions in fact concerned the abrogation of: (1) the prerogative of CIPE to decide the localization of nuclear plants, when the municipalities concerned did not decide; (2) the economic compensation to the municipalities which hosted nuclear or coal plants; and (3) the possibility for ENEL to participate in international nuclear programmes. Nevertheless, almost 80% of the votes in favour of the abrogation evidenced a clear popular will (though the Chernobyl accident undoubtedly played a role). The collaboration to *Superphénix* was immediately suspended, and the project of the PEC abandoned.

Here the last act occured. In the aftermath of the referendum, the Government (the first Goria cabinet) suspended the project of the Trino plant, shut down the Latina plant, and started verifications on the safety of the Caorso plant, and the feasibility of the Montalto di Castro plant under construction (for the non-nuclear parts).

In any case, the final decisions were "Solomonic", as it happens in Italy (one can even suspect that the leading political class seized the moment, being conscious of the insurmountable obstacles met by the nuclear programmes). In fact, in the subsequent years all the Italian nuclear plants were closed, almost every activity in the field of nuclear energy was dismantled, reconverting competencies and agencies to other fields, but leaving a heavy and expensive (although relatively limited) inheritance: four nuclear plants to be decommissioned, an unsafe system of "temporary" radioactive waste storages, the majority of the fuel rods still stored in the deactivation pools, often in precarious conditions (slowly transferred to France for reprocessing).[50]

In 2003, a bungled attempt by one of the Silvio Berlusconi governments to place a national storage site for radioactive wastes at Scanzano Ionico (Basilicata, South Italy), imposing it without any public consultation or information, and declaring a national radioactive emergency, was stopped by a strong uprising of local populations: the national storage site is still waiting, while the nuclear emergency was never called off, and quickly forgotten.

### 14. "Sometimes they come back": the short-lived attempt at a nuclear revival, 25 years later

One would hardly have expected a revival of nuclear ambitions more than 20 years later, "sprung from the hat" after the general election of April 2008, won by Silvio Berlusconi. In brief, the purchase of four not yet tested EPR reactors from French EDF was programmed, negotiations began, a (decidedly nuclear-oriented) National Safety Authority was offhandedly established. In 2010 the IDV party (*Italia dei Valori*) proposed a national referendum in order to stop these programmes. The interest of the Italian population towards the nuclear issue was very dubious, but the vote was called for 11-12 June 2011, together with two more referendums for the nationalization of water services, which were much more felt by the population. Once more, in the community of physicists and technicians the faith in nuclear technology prevailed, although it was weaker than 24 years earlier (and even more sceptical in other scientific milieus). The Democratic Party this time pronounced in favour of the stop to the nuclear programmes, however "tepidly" and inconsistently, since it was against the nationalization of water services. In the meanwhile, on 13 March 2011, the Fukushima earthquake and nuclear disaster happened. This made it easier to form the quorum required in the Italian national referendums, and around 95% of the voters approved the referendum questions, calling for the halt of the nuclear revival. Once more the political decisions were opportunistic: the Safety Agency, which should have been radically transformed, was dissolved instead (while the popular will against privatization of water services was disregarded).

Italy is still left with almost all the problems inherited by its nuclear programmes unsolved. Although modest altogether, these problems are deteriorating and are quite concerning. They have been entrusted to a State joint-stock company, SOGIN (*Società Gestione Impianti Nucleari*), which has been involved in several scandals and they still weigh for around 300-400 million Euros yearly on the bills of Italian consumers. Only now, the programme for a national storage of radioactive wastes is underway, although while we are writing the uncertainties about the acceptance by local populations of the localization that will be proposed lie heavy on it.

**List of acronyms**
AEA, Atomic Energy Agency (UK)
AEC, Atomic Energy Commission (USA)
AGIP, *Azienda Generale Italiana Petroli*
BIRS, International Bank for Reconstruction and Development
CAMEN, *Centro Applicazioni Militari Energia Nucleare*
CEA, "Commissariat pour l'Energie Atomique" (France)
CEE, *Comunità Economica Europea*
CIPE, *Comitato Interministeriale per la Programmazione Economica*
CISAM, *Centro Inter-servizi Studi e Applicazioni Militari*
CISE, *Centro Informazioni Studi Esperienze*
CNR, *Comitato Nazionale per le Ricerche*
CNRN, *Comitato Nazionale per le Ricerche Nucleari*
DC (party), *Democrazia Cristiana*

DISP, *Direzione Sicurezza e Protezione*
EDF, "Electricité de France" (France)
ENEA, *Comitato Nazionale per l'Energia Nucleare*
ENEL, *Ente Nazionale per l'Energia Elettrica*
ENI, *Ente Nazionale Idrocarburi*
ENSI, *Energia Nucleare Sud Italia*
EUREX, Enriched Uranium Extraction
GETSCO, General Electric Technical Services Company
IAEA, Internationl Atomic Energy Agency
IDV (party), *Italia Dei Valori*
IRI, *Istituto per la Ricostruzione Industriale*
ITREC, *Impianto di Trattamento e Rifabbricazione Elementi di Combustibile*
NERSA, *Nucléaire Européenne à Neutron Rapid S.A.*
NPT, Non-Proliferation Treaty
OECD, Organisation for Economic Co-operation and Development
PCUT, *Progetto Ciclo Uranio Torio*
PEC, *Prova Elementi Combustibile*
PCI (party), *Partito Comunista Italiano*
PRO, *Progetto Reattore Organico*
PSDI (party), *Partito SocialDemocratico Italiano*
PUN, Unified Nuclear Project
SELNI, *Società Elettronucleare Italiana*
SIMEA, *Società Italiana Meridionale per l'Energia Atomica*
SENN, *Società ElettroNucleare Nazionale*
SOGIN, *Società Gestione Impianti Nucleari*
SORIN,  *Società Ricerca Impianti Nucleari*

**Notes and references**

## Acknowledgements

The authors are indebted towards Prof. Albert Presas i Puig, of the University "Pompeu Fabra", Barcelona, who involved them in three international workshops concerning nuclear programmes in Europe, which originated this research: "A Comparative Study of European energy Programs", 5-6 December 2008; "A Comparative Study of European energy Programs", 3-4 December 2009; "Going Critical: 70 Years of Nuclear Energy", 5-7 November, 2012.

We are grateful to Dr. Matteo Gerlini, of the University of Florence, for his initial collaboration to this research, and subsequent discussions.

**Disclosure statement**

No potential conflict of interest was reported by the authors.